\documentclass[conference]{IEEEtran}
\IEEEoverridecommandlockouts
\usepackage{cite}
\usepackage{amsmath,amssymb,amsfonts}
\usepackage{algorithmic}
\usepackage{graphicx}
\usepackage{textcomp}
\usepackage{xcolor}
\usepackage{hyperref}
\usepackage{makecell}
\usepackage{subcaption}
\usepackage{titlesec}

\makeatletter
\def\subsection{\@startsection{subsection}{2}{\z@}{1ex plus 0.3ex minus 0.1ex}%
{0.5ex plus 0.1ex}{\normalfont\normalsize\itshape}}

\def\subsubsection{\@startsection{subsubsection}{3}{\parindent}{0.4ex plus 0.1ex minus 0.1ex}
{0ex}{\normalfont\normalsize\itshape}}
\makeatother

\usepackage[font=footnotesize,skip=5pt]{caption}

\def\BibTeX{{\rm B\kern-.05em{\sc i\kern-.025em b}\kern-.08em
    T\kern-.1667em\lower.7ex\hbox{E}\kern-.125emX}}
\begin{document}

\title{
  '1'-bit Count-based Sorting Unit to Reduce Link Power in DNN Accelerators
}
\author{
    \IEEEauthorblockN{Ruichi Han, Yizhi Chen, Tong Lei, Jordi Altayo Gonzalez, Ahmed Hemani}
    \IEEEauthorblockA{
        Department of Electronics and Embedded Systems,
        KTH Royal Institute of Technology, Stockholm, Sweden\\
        \{ruichi, yizhic, tonglei, jordiag, hemani\}@kth.se
    }
}

\maketitle

\begin{abstract}
Interconnect power consumption remains a bottleneck in Deep Neural Network (DNN) accelerators. While ordering data based on '1'-bit counts can mitigate this via reduced switching activity, practical hardware sorting implementations remain underexplored. This work proposes the hardware implementation of a comparison-free sorting unit optimized for Convolutional Neural Networks (CNN). By leveraging approximate computing to group population counts into coarse-grained buckets, our design achieves hardware area reductions while preserving the link power benefits of data reordering.
Our approximate sorting unit achieves up to 35.4\% area reduction while maintaining 19.50\% BT reduction compared to 20.42\% of precise implementation.
 
\end{abstract}

\begin{IEEEkeywords}
‘1’-bit count-based sorting, Approximate computing,  Bit transition reduction, Link power
\end{IEEEkeywords}

\section{Introduction}
\label{sec:intro}

Deep neural network (DNN) accelerators are essential for modern computing~\cite{silvano2025survey}~\cite{samanta2024surveyedge}. While DNN accelerators perform many arithmetic operations, recent works show that data movement dominates power consumption~\cite{armeniakos2022hardware}~\cite{chen2016eyeriss}~\cite{krishnan2021impact}.

On-chip interconnects and NoC fabrics consume a growing share of chip power as models and parallelism scale, particularly in data-intensive workloads~\cite{silvano2025survey}~\cite{krishnan2021impact}. Within NoC-based architectures, physical links account for a significant portion of the energy budget\cite{chen2016eyeriss}~\cite{krishnan2021impact}. As dynamic power is fundamentally driven by the switching activity on links, where each bit transition (BT) incurs capacitive charging costs, reducing BT is an effective lever to improve accelerator efficiency~\cite{armeniakos2022hardware}\cite{chen2016eyeriss}.

Various BT reduction encoding schemes have been developed to suppress interface switching activity~\cite{samanta2024surveyedge}~\cite{jang2025witch}. Standard encoding algorithms often fail to adapt to the unique traffic patterns of neural computing~\cite{silvano2025survey}~\cite{chen2016eyeriss}~\cite{krishnan2021impact}~\cite{jang2025witch}. Recent studies exploit accumulation order-insensitivity, utilizing '1'-bit count or population count (popcount) sorting to minimize switching activity~\cite{chen2025bittransitionreductiondata}. However, these largely theoretical proposals lack practical hardware evaluations to justify the the sorting mechanism's area and power overhead ~\cite{silvano2025survey}~\cite{chen2025bittransitionreductiondata}.

To our knowledge, this is the first hardware implementation of a popcount sorting unit for DNNs with approximate sorting. For terminological consistency, though both terms are equivalent, "popcount" denotes the hardware unit, while "'1'-bit count" for non-hardware discussion. We summarize our contributions as follows:

\begin{enumerate}
    \item \textbf{Hardware Implementation}: 
     Distinct from simulation-based prior work, we present a hardware implementation of comparison-free popcount sorting and compare with other popcount sorting implementations.
    
    \item \textbf{Approximate computing optimization}: 
    We not only implement the accurate sorting unit, but also extend it with approximate sorting to reduce the area cost.
    
    \item \textbf{Comprehensive power analysis}: 
    We analyze ordering's impact on both link power and computational unit power through post-layout power analysis.
    
     \item \textbf{DNN layer's workload}:
     We evaluate our design under a convolution and pooling workload, adopting the same configuration from LeNet's first two layers. 

\end{enumerate}

\textbf{The rest of this paper is organized as follows:} Section~\ref{sec:related} reviews related work. Section~\ref{sec:methodology} describes our comparison-free sorting unit architecture and approximate computing techniques. Section~\ref{sec:results} presents our experiment setup and results. Section~\ref{sec:conclusion} concludes the paper.

\section{Related Work}
\label{sec:related}

The rising power consumption of on-chip interconnects in DNN accelerators drives the need for strategies that minimize dynamic switching costs on links~\cite{krishnan2021impact}. Encoding techniques such as those employed in Eyeriss v2 minimize BT during data transmission through signal-level transformations~\cite{chen2019eyeriss}. However, such encoding approaches introduce encoding/decoding overhead that impacts overall hardware efficiency~\cite{liu2024inspire}.

DNN accumulation operations exhibit order-insensitivity, enabling data reordering strategies to reduce switching activity~\cite{chen2016eyeriss}. Chen \textit{et al.}~\cite{chen2025bittransitionreductiondata} demonstrated that '1'-bit count based ordering reduces link BT but incurs non-trivial hardware overhead. This motivates our focus on minimizing sorting unit area while preserving power benefits~\cite{armeniakos2022hardware}.

Traditional sorting networks like Batcher's Bitonic~\cite{batcher1968bitonic} are comparator-heavy. Constant-time approaches like the Competition Sorter Network (CSN)~\cite{fairouz2025csn}~\cite{jelodari2020o(1)} improve speed but require 80\% more logic elements. Comparison-free architectures offer a more area-efficient foundation by eliminating explicit comparator logic. AdbelHafeez \textit{et al.}~\cite{abdelHafeez2017comparisonfree} proposed a linear-time sorter mapping inputs into hardware-friendly structures, yielding low buffering and minimal complexity. Subsequent refinements include FSM integration for control~\cite{bhargav2019fsmcomparisonfree} and bidirectional data-paths to reduce latency~\cite{chen2021quasicomparisonfree}. OnSort~\cite{gao2025onsort} addressed scalability using sparse-aware metadata filters.

\begin{figure*}[htbp]
    \centering
    \includegraphics[width=1\linewidth]{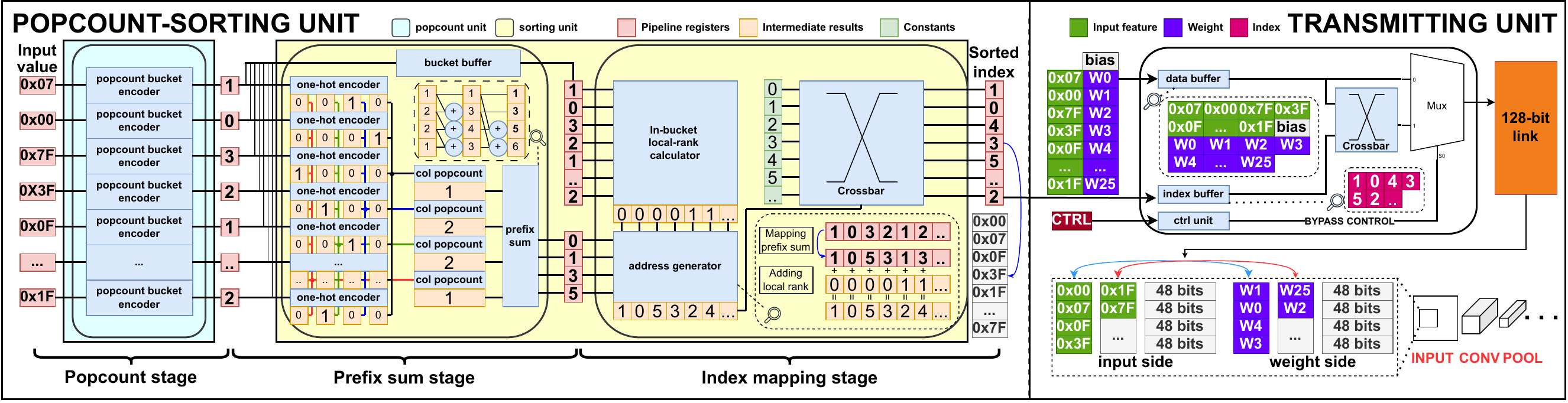}
    \caption{Architecture of the pop-sort unit with an example data flow. }
    \label{fig:sorting_unit_logic}
    \vspace{-8pt}
\end{figure*}
\section{Methodology and Design}
\label{sec:methodology}

Approximate computing improves hardware efficiency across multiple domains. Approximate multipliers~\cite{du2024lowpower} target MAC operations, trading computational accuracy for power reduction, while coarse bucket approximation~\cite{xiao2015approx} targets sorting operations, trading perfect ordering for improved efficiency. The latter approach motivates our application of approximation.

The research goal is to reduce power by minimizing BTs on interconnect links while maintaining a low hardware implementation cost in terms of area overhead.

\subsection{Baseline comparison-free popcount-sorting unit}
The baseline for our work is an Accurate Popcount-Sorting Unit (ACC-PSU) with comparison-free design, adapted from Yang \textit{et al.}~\cite{yang2025explore}. This architecture comprises a popcount unit and a sorting unit, organized into 3 stages: the popcount stage, the prefix sum stage, and the index mapping stage, as illustrated in Fig. \ref{fig:sorting_unit_logic}.  The popcount unit computes the Hamming Weight using 4-bit lookup tables (LUTs) whose outputs are aggregated by adders to produce the '1'-bit count for each input element. The sorting unit encodes '1'-bit count into one-hot representation, generates a frequency histogram, computes starting addresses for each specific '1'-bit count value via prefix sum, and scatters indices into the sorted output. Transmitting units then rearrange input data according to sorted indices and forward data through links.

\subsection{Approximate computing inside popcount-sorting unit}

To reduce hardware overhead, we propose an Approximate Popcount-Sorting Unit (APP-PSU). This design integrates approximate techniques into the population count module. The key innovation lies in grouping exact population counts into coarse-grained buckets, thereby reducing datapath width.

\subsubsection{Effectiveness of approximate sorting}
{Order-insensitive nature of convolution accumulation enables approximate sorting to preserve most benefits of accurate sorting by preventing large transitions, yielding nearly identical BT reduction. Python simulations in Section \ref{sec:results} validate this statement.

\subsubsection{Approximation Design}

The approximation strategy groups exact '1'-bit counts into $k$ coarse-grained buckets through a deterministic mapping. For $W$-bit inputs, each exact '1'-bit count value $P_{exact}\in[0,W]$ is assigned to one of $k$ buckets, producing a compact bucket index requiring only $log_2(k)$ bits (e.g., a 2-bit index for $k=4$ buckets). Rather than preserving the full '1'-bit count range, only these bucket indices are forwarded to the sorting stage, compressing the value space from $W+1$ possible '1'-bit counts to $k$ buckets.

For example, for $W=8$-bit inputs and $k=4$ buckets, the mapping assigns $\{0,1,2\} \rightarrow Bucket 0$, $\{3,4\}\rightarrow Bucket 1$, $\{5,6\}\rightarrow Bucket 2$, and $\{7,8\}\rightarrow Bucket 3$. An input sequence with exact '1'-bit counts $\{4,1,7,5,3,5\}$ would thus map to bucket indices $\{1,0,3,2,1,2\}$.

\subsubsection{Hardware Implementation and Architecture}

The APP-SU architecture, as in Fig.\ref{fig:sorting_unit_logic}, retains the same dataflow as ACC-PSU while introducing a mapping LUT in popcount bucket encoder for approximation. During synthesis, the compiler eliminates logic paths that do not affect the final bucket index, so the synthesized netlist is a simplified approximate '1'-bit count circuit that directly produces bucket indices.

The primary area reduction comes from reducing the number of buckets, which determines the datapath width throughout the sorting unit -- all sorting logic scales proportionally with the bucket count rather than the full '1'-bit count range. For $W=8$-bit data, ACC-PSU requires $W+1=9$ buckets to cover every possible '1'-bit count value, whereas APP-PSU aggregates adjacent '1'-bit count values into $k=4$ buckets.

\section{Experiment}
\label{sec:results}

\begin{figure*}[htbp]
    \centering
    \includegraphics[width=1\linewidth]{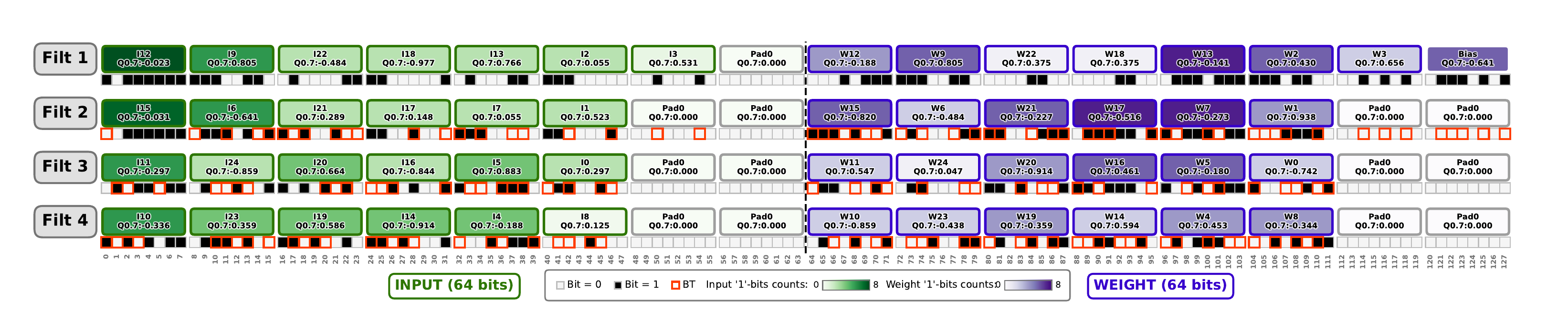}
    \caption{Examples of ordered on 128-bit links after our APP-PSU.}
    \label{fig:bit_flip_order}
    \vspace{-8pt}
\end{figure*}

We evaluate four configurations: a non-optimized baseline, column-major ordering, ACC ordering, and APP ordering with $k = 4$ buckets. All experiments use 8-bit fixed-point representation. We also compare our PSUs with other sorting implementations. For the hardware evaluation, ACC-PSU and APP-PSU are implemented on our platform, whereas the baseline is implemented as a bypass path.
\subsection{Results of Software Simulation Experiment }

\begin{table}[htbp]
\vspace{5pt}
\caption{Bit flip under different order strategy}
\vspace{-8pt}
\begin{center}
\begin{tabular}{|l|c|c|c|c|}
\hline
\multicolumn{1}{|c|}{\textbf{Order}}&\multicolumn{4}{c|}{\textbf{Bit Transitions per 128-bit flit}} \\
\cline{2-5} 
\multicolumn{1}{|c|}{\textbf{strategy}} & \textbf{\textit{Input}}& \textbf{\textit{Weight}}& \textbf{\textit{Overall}}& \textbf{\textit{Reduction}} \\
\hline
Non-optimized       & 31.035 & 32.036 & 63.072 & - \\
\hline
Column-major     & 26.004 & 28.007 & 54.011 & 14.366\% \\
\hline
ACC Ordering   & 22.333 & 28.013 & 50.346 & 20.177\% \\
\hline
\textbf{APP Ordering}   & 22.887 & 28.009 & 50.896 & \textbf{19.305\%} \\
\hline
\end{tabular}
\label{tab1}
\end{center}
\vspace{-5pt}
\end{table}

We simulate packet transfers over a 128-bit link using Python. A total of 100000 packets are sent, each consisting of 4 flits with random inputs and weights. As shown in Table \ref{tab1}, adopting column-major ordering reduces overall BT by 14.366\% compared with the non-optimized baseline. Furthermore, when applying our ordering strategies based on the input '1'-bit count, the input-side BT decreases further, leading to overall BT reductions of 20.177\% and 19.305\% for ACC ordering and APP ordering, respectively.

\subsection{Results of DNN Workload Experiment }
Fig.~\ref{fig:platform} shows our evaluation platform, which consists of a data allocation unit and 16 processing elements (PEs) implementing the first convolution and pooling layers of LeNet-5. The allocation unit supplies each PE with inputs, weights, and bias. Inside the allocation unit, the sorting unit generates sorted indices, and the transmitting units permute the data according to these indices before forwarding them to PEs.
\begin{figure}[htbp]
    \centering
    \includegraphics[width=1\linewidth]{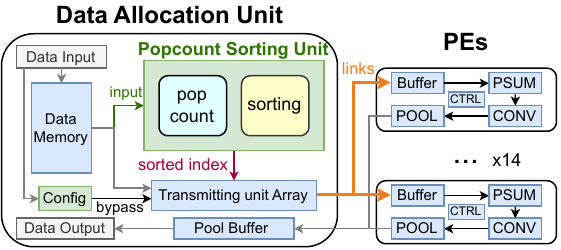}
    \caption{Platform with sorting units and PEs for convolution and pooling layers.}
    \label{fig:platform}
\end{figure}
\subsubsection{Waveform verification of APP-PSU}
We present the waveform of four representative input patterns to illustrate the behavior of the APP-PSU. As shown in Fig.~\ref{fig:questasim_ordering}, output indices from buckets with higher ‘1’-bit counts are placed after those from buckets with lower counts. For (1) all-ones and (2) all-zeros patterns, the output indices follow ascending order. In the case of (3) repeated pattern whose '1'-bit count decreases from 8 to 0, the last eight indices shown in the figure clearly illustrate the behavior of the APP ordering algorithm. The same ordering effect is observed for (4) random input pattern.

\begin{figure}[htbp]
    \includegraphics[width=1\linewidth]{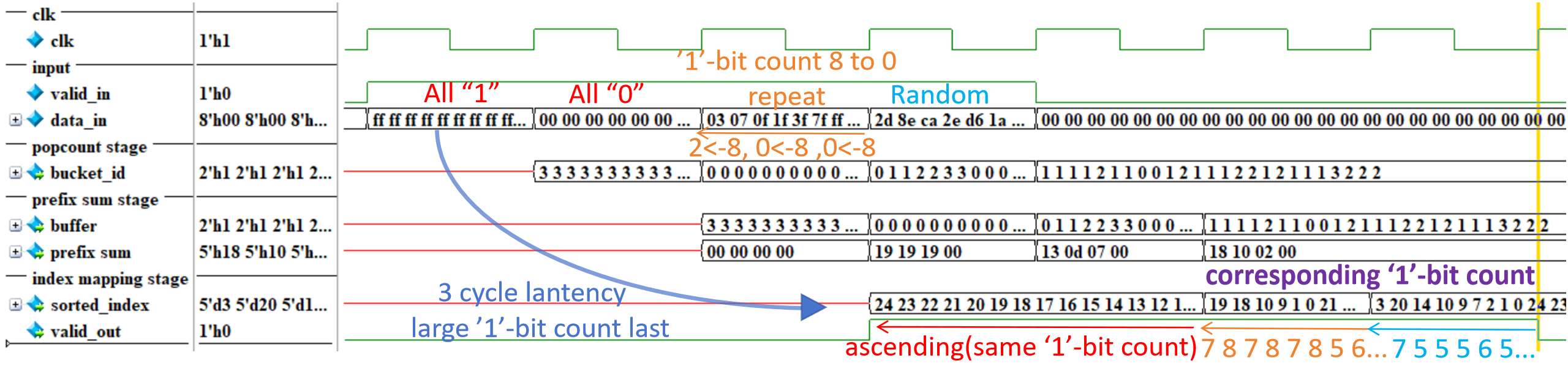}
    \caption{QuestaSim waveform of APP-PSU.}
    \label{fig:questasim_ordering}
\end{figure}

\subsubsection{Example of ordered flits}
Fig.~\ref{fig:bit_flip_order} shows a 128-bit link transmission snapshot for a single packet processed by the APP-PSU. On the input side, the ‘1’-bit counts of the transmitted values exhibit a generally decreasing trend; for example, data with a '1'-bit count of 7 are placed ahead of those with counts of 5 or 6, whereas the weight side remains random. This ordering reduces abrupt large popcount variations and helps maintain a small BT gradient across the link.

\subsubsection{Sorting unit area reduction}
We synthesized four popcount-sorting unit designs -- one based on the Bitonic network~\cite{batcher1968bitonic}, one based on CSN~\cite{fairouz2025csn}~\cite{jelodari2020o(1)}, ACC-PSU, and APP-PSU -- using commercial EDA tools targeting 500 MHz in a 22 nm technology node, with all designs implemented using the same pipeline depth. The evaluation covered two convolution kernel sizes, $5\times5$ and $7\times7$. As shown in Fig.~\ref{fig:area_comp}, APP-PSU attains the lowest area of $2193\mu m^2$ and $6928\mu m^2$ for the two kernel sizes, outperforming all other architectures. The approximation yields 24.9\% and 36.7\% area reductions in the popcount unit and sorting unit, respectively, at a kernel size of 25, resulting in an overall reduction of 35.4\%.

\begin{figure}[htbp]
    \centering
    \includegraphics[width=1\linewidth]{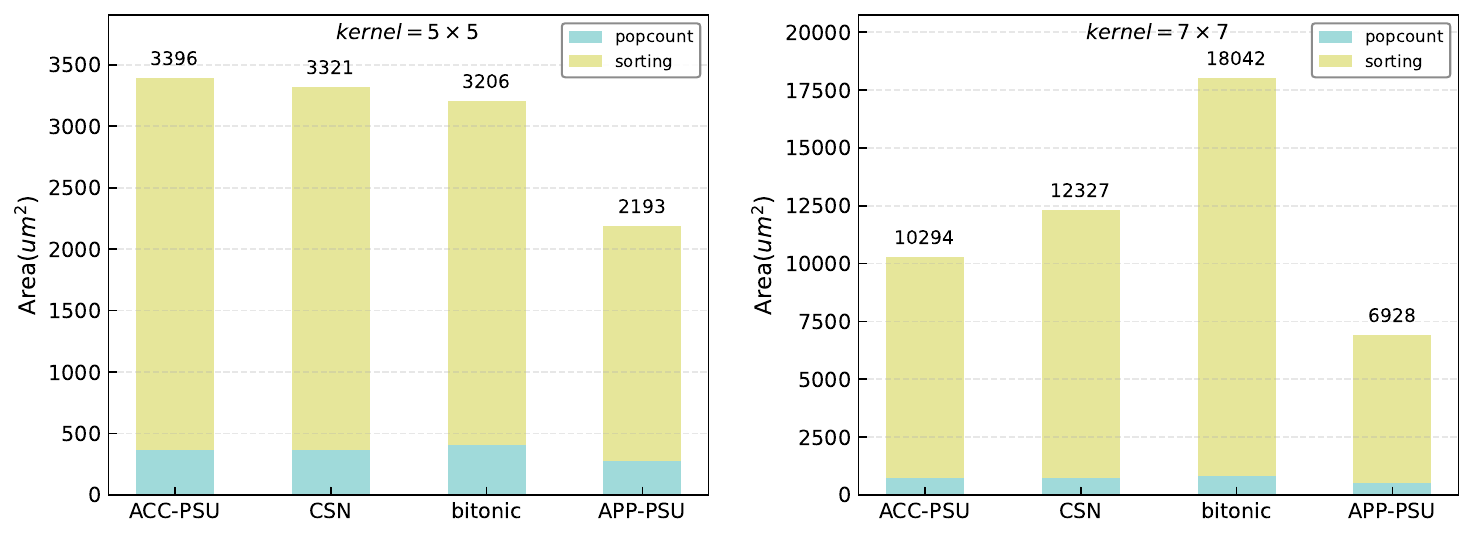}
    \caption{Area breakdown of different Sorting Unit designs. }  
    \label{fig:area_comp}
\end{figure}

\subsubsection{Power result}

We evaluate post-layout power consumption with back-annotated switching activity usinc commercial EDA tools. A set of 100 convolution kernels is applied as test vectors to three configurations: the non-optimized baseline, ACC ordering, and APP ordering. All reported percentages are averaged. At the PE level, ACC-PSU reduces power by 4.98\%, while APP-PSU achieves a 4.58\% reduction. The corresponding PE power breakdown for APP ordering versus baseline is shown in Fig.~\ref{fig:power_pie_chart}. 

\begin{figure}[htb]
    \centering
    \includegraphics[width=0.8\linewidth]{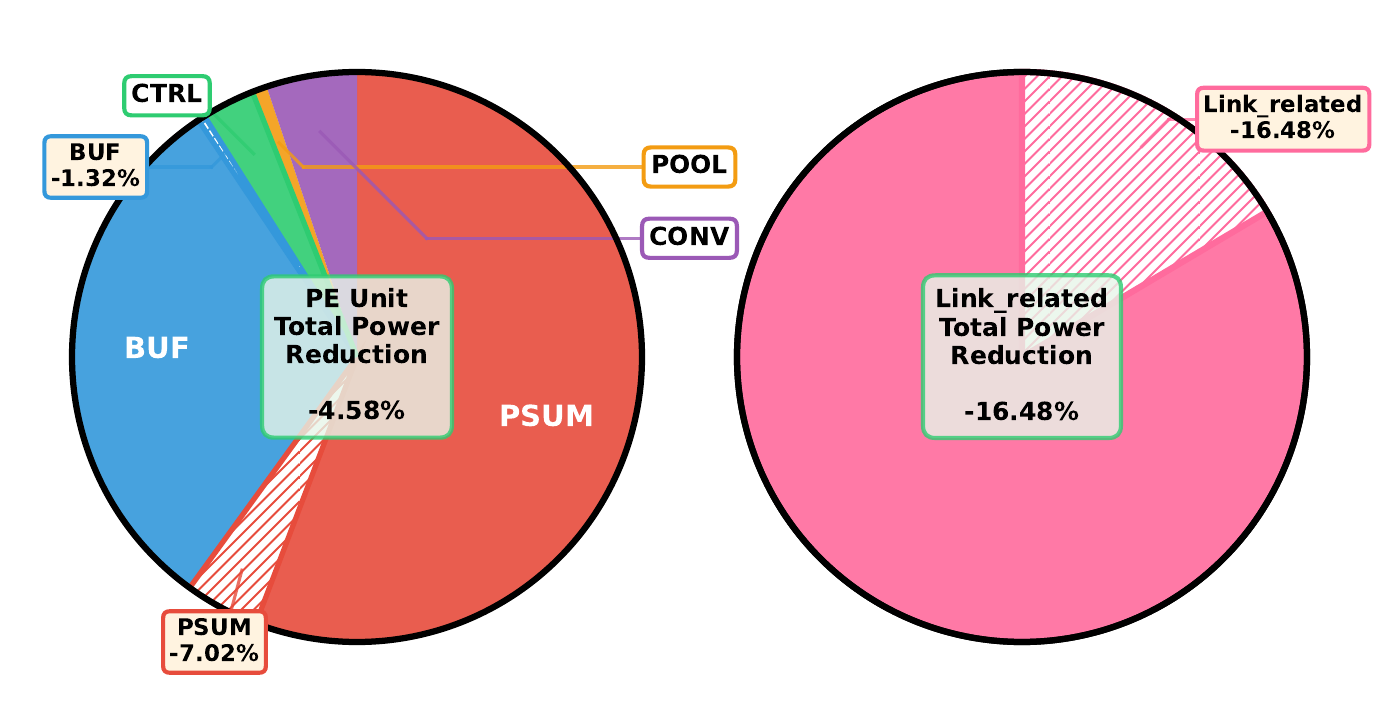}
    \caption{ Breakdown of achieved power reduction in non-link and link power.}
    \label{fig:power_pie_chart}
\end{figure}

\begin{figure*}[htb]
    \centering
    \includegraphics[width=1\linewidth]{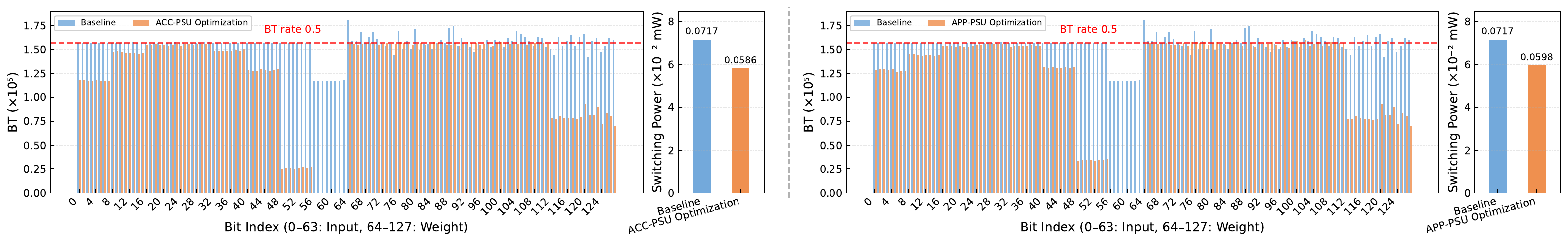}
    \caption{BT reduction and link-related power reduction (ACC-PSU, left; APP-PSU, right).}
    \label{fig:link_comp}
    \vspace{-8pt}
\end{figure*}

The switching power of the transmission registers is extracted as a proxy for link power, referred to as link-related power. As shown in Fig.~\ref{fig:link_comp}, ACC-PSU reduces link-related power by 18.27\% and link BT by 20.42\%, while APP-PSU achieves 16.48\% and 19.50\%, respectively.

Although APP-PSU yields slightly smaller power savings than ACC-PSU, its power overhead is only 1.43 mW compared to ACC-PSU's 2.28 mW, corresponding to a 37.3\% reduction.

\subsection{Discussion and limitation}

\textbf{1. BT Reduction Effectiveness} Our proposed APP-PSU achieves a 19.50\% BT reduction in links compared with the non-optimized baseline, retaining 95.5\% of the 20.42\% BT reduction provided by ACC-PSU.

\textbf{2. Hardware Efficiency of APP-PSU} APP-PSU reduces link-related power by 16.48\%, corresponding to 90.2\% of the link-power savings achieved by ACC-PSU. At the same time, it lowers area overhead by 35.4\% and power overhead by 37.3\%, resulting in a substantially lower hardware cost.

\textbf{3. End-to-end power and hop count} Our simplified platform uses only single-hop traversal for each packet, achieving a 16.48\% link-power reduction with a 1.43~mW overhead. Our methods should achieve scalably greater savings with more hops in a full-scale accelerator with longer, multi-hop interconnects, as BT reduction benefits accumulate at each router-to-router transmission. Since this study focuses on the sorting unit, the platform serves as an illustrative example.

\textbf{4. Limitation} As future work, we plan to extend the analysis from a single convolution layer to more complex neural networks, such as ResNets and Transformers and to evaluate BT reduction and power saving over multi-hop paths within an accelerator platform with a detailed NoC implementation. 

\section{Conclusion}
\label{sec:conclusion}

We present the hardware implementation of a popcount sorting unit for partial LeNet. By introducing approximate computing to simplify logic, both area and power are reduced. Post-layout analysis in 22 nm technology demonstrates that our APP-PSU achieves 35.4\% area reduction and 37.3\% power reduction compared to the accurate baseline, while maintaining 95.5\% BT reduction efficiency (19.5\% vs 20.4\%). Post-layout power analysis with back-annotated switching activity confirms 16.48\% link-related power savings and 4.58\% PE power reduction. 
Future work will scale to larger models (e.g., ResNets, Transformers) and integrate our method into complete NoC-based platforms to quantify end-to-end benefits

\bibliographystyle{IEEEtran}
\bibliography{ref1}
\end{document}